\documentclass[showpacs,showkeys,pre]{revtex4-1}
\usepackage{amsmath}
\usepackage{hyperref}
\usepackage{graphicx}

\usepackage{natbib} 			
\usepackage{amsfonts}
\usepackage{amsmath}
\usepackage{amssymb}
 \usepackage{amsthm}
\usepackage{booktabs}
\usepackage{epsfig}

\begin{document}
\title{Detrended fluctuation analysis as a regression framework: Estimating dependence at different scales}
%\title{Regression framework based on detrended fluctuation analysis: Estimating the dependence at different scales}
\author{Ladislav Kristoufek\footnote{Institute of Information Theory and Automation, Academy of Sciences of the Czech Republic, CZ-182 08, E-mail: kristouf@utia.cas.cz}\footnote{Warwick Business School, University of Warwick, Coventry, West Midlands, CV4 7AL, United Kingdom}
}
%\date{\today}

\begin{abstract}
We propose a framework combining detrended fluctuation analysis with standard regression methodology. The method is built on detrended variances and covariances and it is designed to estimate regression parameters at different scales and under potential non-stationarity and power-law correlations. The former feature allows for distinguishing between effects for a pair of variables from different temporal perspectives. The latter ones make the method a significant improvement over the standard least squares estimation. Theoretical claims are supported by Monte Carlo simulations. The method is then applied on selected examples from physics, finance, environmental science and epidemiology. For most of the studied cases, the relationship between variables of interest varies strongly across scales. 
\end{abstract}


\maketitle

\section{Introduction}

Detrended fluctuation analysis (DFA) was introduced in early 1990s \cite{Peng1993,Peng1994,Peng1995} as a method for analyzing fractal properties of underlying data. The method was later majorly popularized in the long-range correlations \cite{Buldyrev1995,Kantelhardt2001} and multi-fractal analyses \cite{Kantelhardt2002}. Recently, DFA has been generalized for long-range cross-correlations analysis \cite{Podobnik2008,Podobnik2009a,Zhou2008,Jiang2011} as well as an examination of correlations between non-stationary series \cite{Zebende2011,Kristoufek2014}. The method has been applied and utilized across wide range of disciplines ranging from physiology to cardiology, DNA analysis and neurology to (hydro)meteorology, economics and finance, engineering and environment and many others \cite{Stanley1999,Bunde2000,Peng1995a,Buldyrev1998,Montez2009,Talkner2000,Liu1997,Varotsos2002,Weber2001}. Here, we propose a framework based on the detrended fluctuation analysis which allows for a regression analysis of possibly non-stationary and long-range dependent data at different scales.

The methodology is based on the least squares framework, which is shortly recalled and translated into the language of variances and covariances. The detrended fluctuation analysis together with its bivariate generalization of the detrended cross-correlation analysis (DCCA) \cite{Podobnik2008} are described in some detail as a connecting bridge to the DFA-based regression. The DFA framework is introduced for the bivariate setting with procedures to estimate parameters, standard errors of the estimates and the coefficient of determination ($R^2$) all characteristic for a specified scale. The theoretical concepts are further supported by Monte Carlo simulations. The framework is then applied on several phenomena from various disciplines -- relationship between temperature and humidity, stock market betas, elasticity between corn and ethanol, and transmission between influenza outbursts and the Google Flu Trends indicator. For three out of four cases, we report a strong variability of the estimates across scales. The proposed methodology thus provides a further step in the development of DFA and related methods, here specifically from the correlation to regression framework.

\section{Methodology}
\subsection{Least squares regression framework}

When studying dependence between two series, one often considers a linear model in its simplest form
\begin{equation}
Y=\alpha+X\beta+u \nonumber
\end{equation}
where $Y$ is a dependent (response) variable, $X$ is an independent (impulse) variable, $u$ is an error-term, and parameters $\alpha$ and $\beta$ represent relationship between $X$ and $Y$. Estimation of parameter $\beta$ then becomes a crucial point of empirical studies across disciplines. Contrary to the frequently used correlation coefficients (e.g. Pearson, Spearman and Kendall correlation coefficients), $\beta$ is not normalized so that it displays an actual effect of variable $X$ on variable $Y$. The standard regression analysis uses the (ordinary) least squares method for estimation of the parameter of interest $\beta$ as
\begin{equation}
\label{beta}
\widehat{\beta}^{LS}=\frac{\sum_{t=1}^T{\left(x_t-\bar{x}\right)\left(y_t-\bar{y}\right)}}{\sum_{t=1}^T{\left(x_t-\bar{x}\right)^2}}\sim \frac{\widehat{\sigma_{XY}}}{\widehat{\sigma_X}^2},
\end{equation}
where $\bar{x}=\frac{1}{T}\sum_{t=1}^T{x_t}$ and $\bar{y}=\frac{1}{T}\sum_{t=1}^T{y_t}$. Variance of the estimator is obtained using the residuals $\widehat{u}_t=y_t-x_t\widehat{\beta}^{LS}$ as
\begin{equation}
\label{sigma}
\text{var}\left(\widehat{\beta}^{LS}\right)=\frac{\frac{\sum_{t=1}^T{\widehat{u}_t^2}}{T-2}}{\sum_{t=1}^T{\left(x_t-\bar{x}\right)^2}}\sim\frac{1}{T-2}\frac{\widehat{\sigma_u}^2}{\widehat{\sigma_X}^2}
\end{equation}
and it illustrates accuracy of the estimated parameter. Variance can be further utilized in the hypothesis testing. To describe quality of the model, the coefficient of determination $R^2$ defined as
\begin{equation}
\label{R2}
R^2=1-\frac{\sum_{t=1}^T{\widehat{u}_t^2}}{\sum_{t=1}^T{\left(y_t-\bar{y}\right)^2}}\sim 1-\frac{\widehat{\sigma_u}^2}{\widehat{\sigma_Y}^2}
\end{equation}
and ranging between 0 and 1 is utilized. $R^2$ quantifies a proportion of variance of $Y$ explained by $X$ and thus a higher value of $R^2$ signifies a better information content of $X$ in explaining $Y$. On the right-hand side of Eqs. \ref{beta}-\ref{R2}, we translate the standard notation into estimated variances and covariances using the $\widehat{\sigma}$ notation. Evidently, the whole framework is based on estimated variances of $X$, $Y$ and $u$ and covariance between $X$ and $Y$. We employ the same idea using the detrended fluctuation analysis methodology, which we now shortly recall.

\subsection{Detrended fluctuation and cross-correlation analyses}

For time series $x_t$, we construct a profile as $X_t=\sum_{i=1}^t{(x_i-\bar{x})}$ which is split into non-overlapping boxes of length (scale) $s$. In each box between $j$ and $j+s-1$, the linear (or in practice any other) fit of a time trend $\widehat{X_{k,j}}$ is constructed for $j\le k \le j+s-1$. Fluctuation function $f^2_X(s,j)$ is then defined for each box of length $s$ as
\begin{equation}
f_{X}^2(s,j)=\frac{\sum_{k=j}^{j+s-1}{(X_k-\widehat{X_{k,j}})^2}}{s-1}. \nonumber
\label{DFA1}
\end{equation}
The fluctuation $f_X^2(s,j)$ is further averaged over all boxes of length $s$ to obtain
\begin{equation}
F_{X}^2(s)=\frac{\sum_{j=1}^{T-s+1}{f_{X}^2(s,j)}}{T-s}.
\label{DFA2}
\end{equation}
For bivariate series $x_t$ and $y_t$, the procedure is parallel and we get
\begin{equation}
f_{XY}^2(s,j)=\frac{\sum_{k=j}^{j+s-1}{(X_k-\widehat{X_{k,j}})(Y_k-\widehat{Y_{k,j}})}}{s-1} \nonumber
\label{DCCA1}
\end{equation}
which is again averaged over all boxes %\footnote{For computational efficiency, we use non-overlapping boxes of size $s$. In the cases when $T/s$ is not an integer, we average the fluctuations of boxes split both from the beginning and the end of the series so that we obtain $2\lfloor T/s\rfloor$ boxes.} 
with scale $s$ to obtain
\begin{equation}
F_{XY}^2(s)=\frac{\sum_{j=1}^{T-s+1}{f_{XY}^2(s,j)}}{T-s}.
\label{DCCA2}
\end{equation}
The scale-characteristic fluctuations $F_X^2(s)$ and $F_{XY}^2(s)$ can be seen as scale-dependent variance and covariance, respectively. We thus stop in the middle of the DFA and DCCA procedures as our ultimate goal is not to obtain the scaling exponents but only the detrended variances and covariances.

\subsection{DFA-based regression}

We now utilize such correspondence to reformulate the standard regression framework. Estimator in Eq. \ref{beta} can be rewritten for a given scale $s$ as
\begin{equation}
\widehat{\beta}^{DFA}(s)=\frac{F_{XY}^2(s)}{F_{X}^2(s)}
\end{equation}
with a use of fluctuations defined in Eqs. \ref{DFA2}-\ref{DCCA2}. Using the estimated $\widehat{\beta}^{DFA}(s)$, we obtain scale-specific residuals as
\begin{equation}
\widehat{u}_t(s)=y_t-x_t\widehat{\beta}^{DFA}(s)-\overline{y_t-x_t\widehat{\beta}^{DFA}(s)} \nonumber
\end{equation}
with a mean value of zero. These are further plugged into the DFA procedure so that the fluctuation $F_{u}^2(s)$ can be used for estimating variance of $\widehat{\beta}^{DFA}(s)$ via Eq. \ref{sigma} as 
\begin{equation}
\text{var}\left(\widehat{\beta}^{DFA}(s)\right)=\frac{1}{T-2}\frac{F_{u}^2(s)}{F_{X}^2(s)}.
\end{equation}
Eq. \ref{R2} is then translated into the DFA framework as
\begin{equation}
R^2(s)=1-\frac{F_u^2(s)}{F_Y^2(s)}.
\end{equation}
The whole standard regression framework in Eqs. \ref{beta}-\ref{R2} is thus transferred into a scale-dependent framework using the DFA methodology. Moreover, DFA provides some desirable statistical properties such as resistance to non-stationarity and trends which further enhance the proposed methodology \cite{Kantelhardt2001,Kantelhardt2002,Podobnik2008,Kristoufek2014}.

\begin{figure}[htbp]
\begin{center}
\begin{tabular}{c}
\includegraphics[width=80mm]{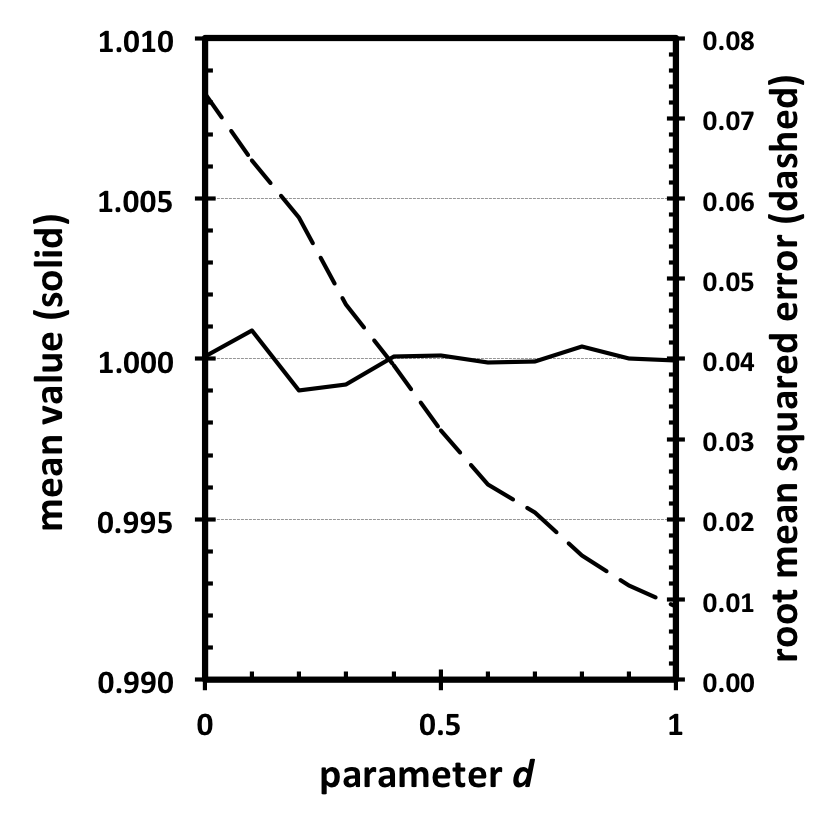}
\end{tabular}
\end{center}
\caption{\textbf{Simulations results I.} {Mean values (solid line, left axis) and root mean squared errors (dashed line, right axis) are shown for model $y_t=\alpha+x_t\beta+u_t$ with $\alpha=\beta=1$, $x_t$ is defined as an ARFIMA process with changing parameter $d_x$ ($x$-axis) and $u_t$ is a standard Gaussian noise error-term. Each simulation has 1000 observations and 1000 series are generated for each setting. Results show that the DFA estimator of $\beta$ is unbiased (mean values range between 0.999 and 1.001) and its variance decreases with the memory strength.}\label{fig_sim1}
}
\end{figure}

\begin{figure}[!htbp]
\begin{center}
\begin{tabular}{c}
\includegraphics[width=80mm]{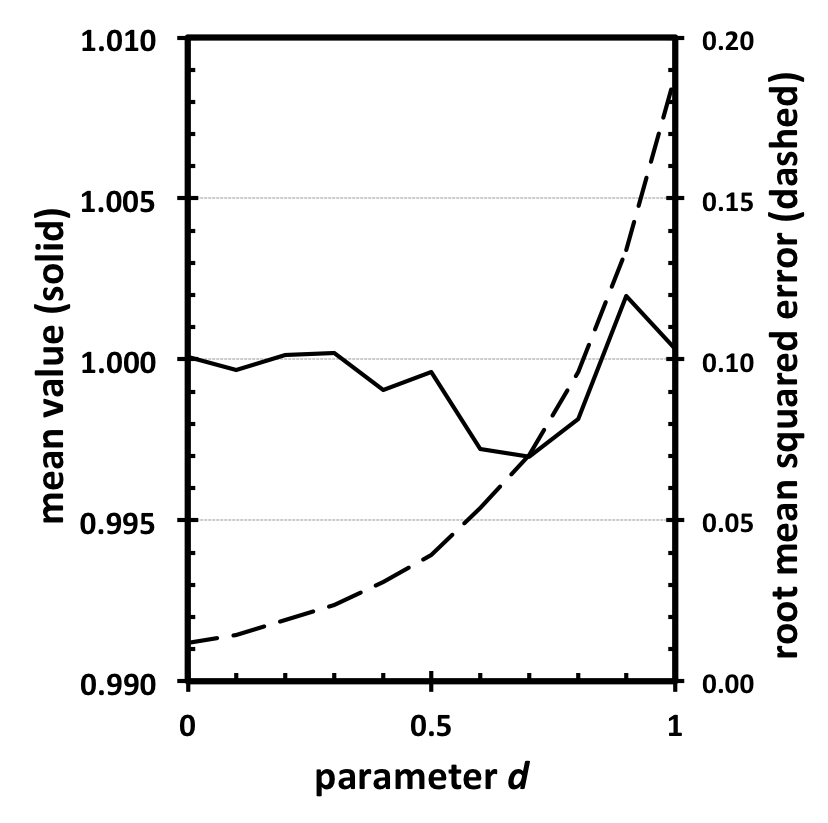}
\end{tabular}
\end{center}
\caption{\textbf{Simulations results II.} {Mean values (solid line, left axis) and root mean squared errors (dashed line, right axis) are shown for model $y_t=\alpha+x_t\beta+u_t$ with $\alpha=\beta=1$, $x_t$ is defined as an ARFIMA process with fixed $d_x=0.9$ and $u_t$ is an ARFIMA process with changing parameter $d_u$ ($x$-axis). Each simulation has 1000 observations and 1000 series are generated for each setting. Results show that the DFA estimator of $\beta$ is still unbiased (mean values range between 0.997 and 1.002) while variance increases with the error-term memory strength.}\label{fig_sim2}
}
\end{figure}

In order to examine performance of the estimator, we study its properties in two non-stationary regression frameworks with $y_t=\alpha+ x_t\beta+u_t$. Firstly, we show how the DFA estimator performs under various levels of long-range dependence in series $x_t$ and $y_t$. The former series is simulated as an ARFIMA process so that $x_t=\sum_{i=1}^{+\infty}{a_i(d)x_{t-i}}$ where $d$ is a fractional integration parameter and $a_i(d)=\frac{\Gamma(i-d)}{\Gamma(-d)\Gamma(1+i)}$. Error-term $u_t$ is taken as a standard Gaussian noise so that the series $y_t$ has the same parameter $d$ as the series $x_t$. Fig. \ref{fig_sim1} shows the mean values and root mean squared errors of the DFA estimator for the series of length 1000 with parameter $d$ ranging between 0 and 1 with a step of 0.1. The estimator is averaged over scales between 10 and 100 with a step of 10, and the regression parameters are set to $\alpha=\beta=1$. 1000 simulations are run for each setting. The estimator is unbiased regardless of the long-range dependence level. Moreover, the root mean squared error decreases with an increasing memory which is desirable. And secondly, we study how the estimator fares for long-range dependent error-terms $u_t$. To do so, we fix the memory parameter for the series $x_t$ to $d_x=0.9$ and the error-term is generated as an ARFIMA process with $d_u$ ranging between 0 and 1 with a step of 0.1. The rest of the setting remains unchanged. In Fig. \ref{fig_sim2}, we again report the mean values and root mean squared errors of the estimator as for the previous case. The DFA estimator is again remarkably stable and unbiased for different levels of memory in the error-terms. Even though variance of the estimator increases with $d_u$, which is expected due to an increasing weight of the error-term in the whole dynamics of $y_t$ as variance of the error-term increases with $d_u$, the overall performance remains excellent.

\section{Applications and discussion}

\begin{figure}[!htbp]
\begin{center}
\begin{tabular}{c}
\includegraphics[width=80mm]{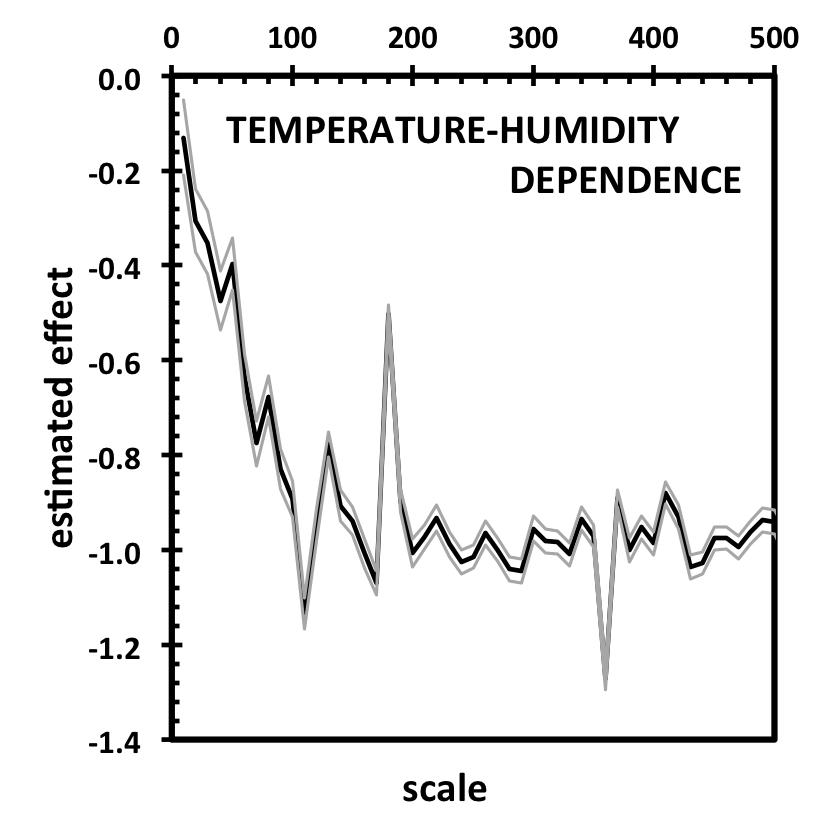}
\end{tabular}
\end{center}
\caption{\textbf{Dependence between air temperature and relative humidity.} {Dependence between the two variables shows a strong variability across scales. The effect is weak for low scales but strengthens for higher scales. For medium and high scales, we find a one-to-one correspondence between changes in temperature and humidity. Scale dependent estimates are represented by a black curve and 95\% confidence intervals are shown using the grey curves.}\label{fig_air}
}
\end{figure}

We utilize the method to analyze four datasets from various disciplines, which usually utilize detrended fluctuation analysis and related methods. First, we study a relationship between daily air temperature and relative humidity in London, UK, between years 2000 and 2012 (4651 observations). An increasing temperature is expected to increase a moisture holding capacity of air and thus decrease its relative humidity. Fig. \ref{fig_air} shows an estimated effect of temperature on relative humidity for scales between 10 and 1150 (approximately a quarter of the time series length) days. Expectedly, the effect is negative. However, a strong variation across scales is uncovered. The effect is weak for low scales but its strength increases for higher scales. After approximately quarter of a year, the effect reaches values around unity and even though some further variation is visible, the effect remains fairly close to one. The effect is thus found to be rather cumulative than instantaneous and it takes several months before the temperature changes fully translate into relative humidity. An increase of one degree Celsius in mean daily temperature is accompanied by an eventual decrease of one percentage point in relative humidity. Narrow confidence intervals suggest high reliability of the estimates.  %etc\

\begin{figure}[!htbp]
\begin{center}
\begin{tabular}{c}
\includegraphics[width=80mm]{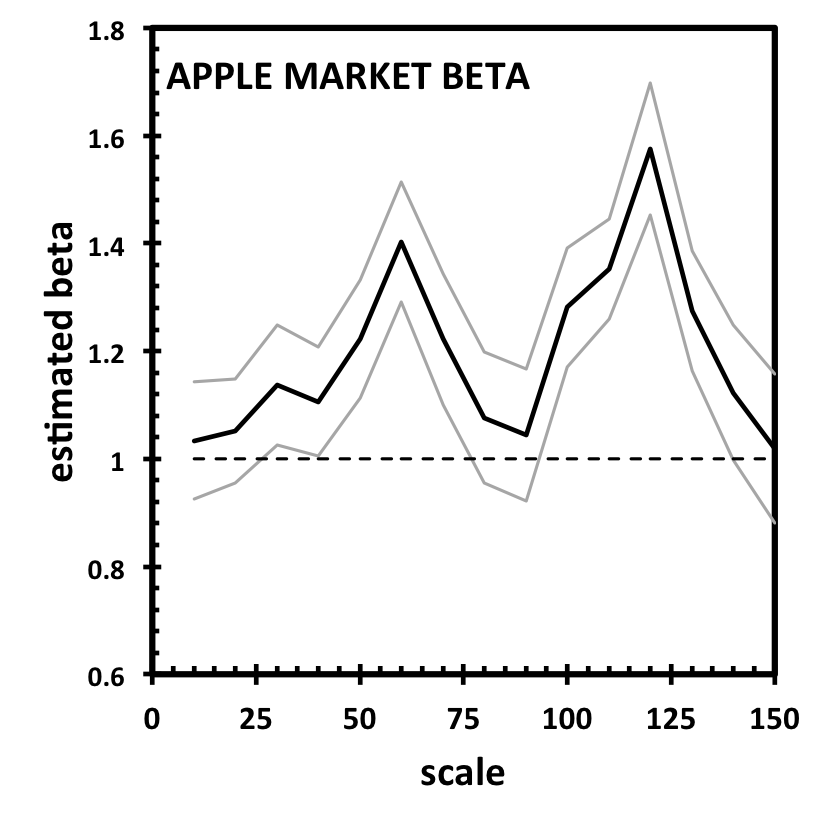}\\
\includegraphics[width=80mm]{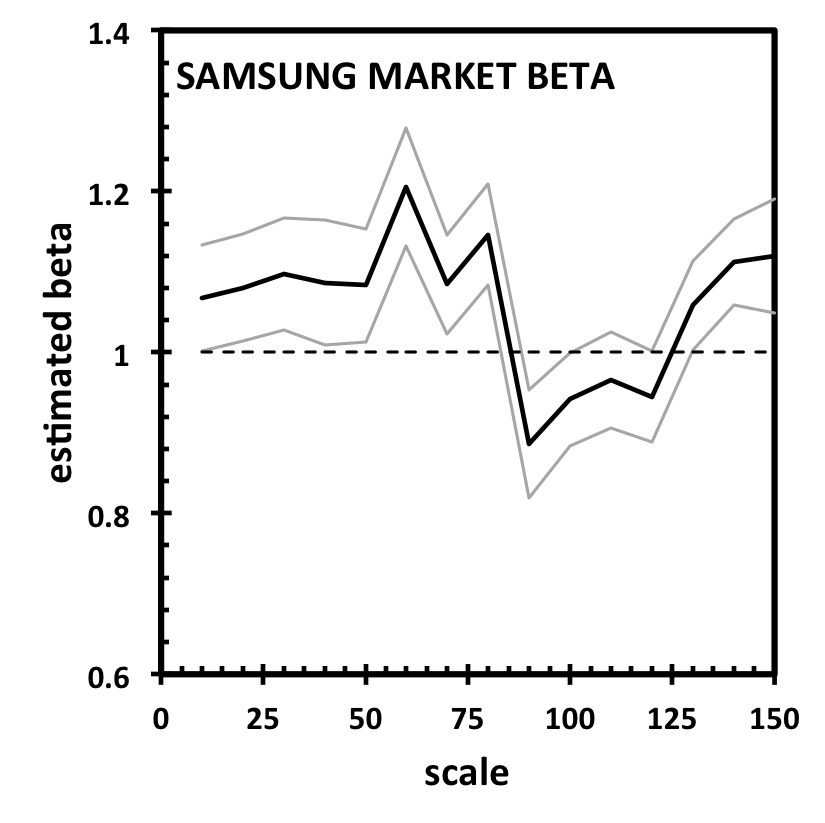}
\end{tabular}
\end{center}
\caption{\textbf{Systematic risk of Apple and Samsung stocks.} {Market betas are estimated for Apple (top) and Samsung (bottom) stocks relative to their respective market indices (NASDAQ and KOSPI, respectively). Estimated sensitivity of the stocks to their indices is quite stable across scales and remains close to one. Scale-dependent estimates are represented by a black curve and 95\% confidence intervals are shown using the grey curves. A black dashed line represents a hypothetical market-following stock with $\beta=1$.}\label{fig_CAPM}
}
\end{figure}

Second, we analyze a relationship between stock returns and related stock indices, which forms a building block of the capital asset pricing model (CAPM) in finance. We focus on two competing companies and their stocks -- Apple, Inc. (traded on NASDAQ, USA) and Samsung Electronics Co. Ltd. (traded on KOSPI, South Korea). The CAPM relationship stems in a simple model $r_{i,t}=\alpha+\beta r_{M,t}+u_t$ where $r_{i,t}$ and $r_{M,t}$ are returns of a stock and a relevant stock index, respectively. Both parameters -- $\alpha$ and $\beta$ -- have important implications in classical financial economics. The former one is standardly taken as a measure of under- or over-pricing of the stock, and the latter one is a measure of systematic risk due to overall market conditions. We study the systematic risk (market betas) of Apple and Samsung stocks with respect to their respective markets on weekly data between January 2000 and August 2014 (764 observations). Fig. \ref{fig_CAPM} uncovers that both stocks are very tightly connected to their market indices with $\beta$ very close to one. Apple stocks seem to be more aggressive than Samsung stocks as the former $\beta$ remains above one more frequently. Variation of the systematic risk parameters across scales is much weaker than for the previous case and confidence intervals are much wider mainly due to high volatility of financial returns.

\begin{figure}[!htbp]
\begin{center}
\begin{tabular}{c}
\includegraphics[width=80mm]{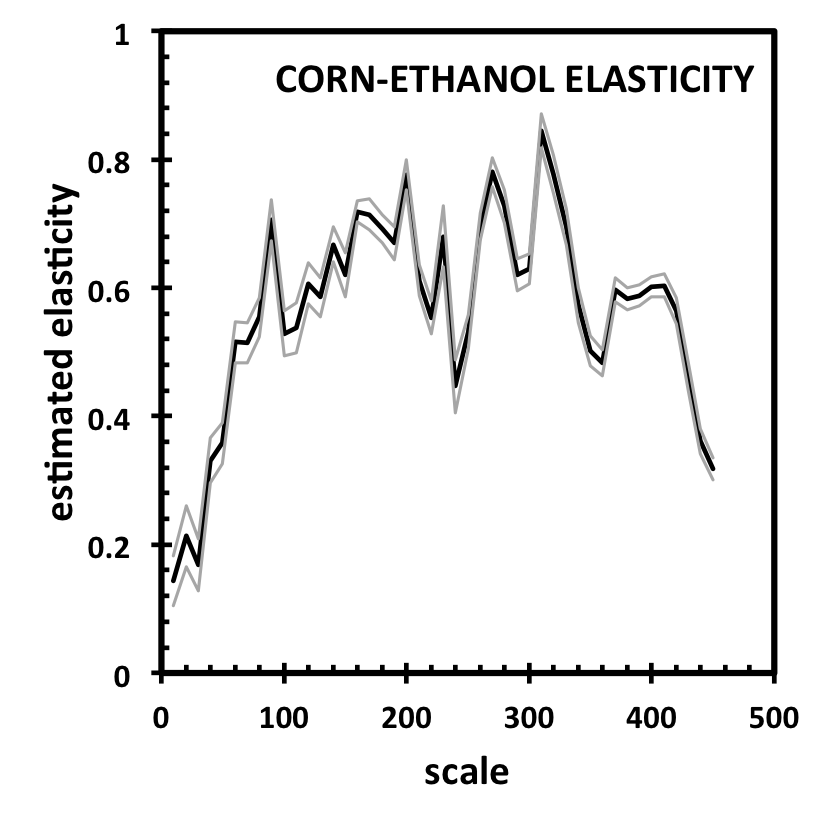}
\end{tabular}
\end{center}
\caption{\textbf{Elasticity between ethanol and corn.} {Estimated elasticity varies considerably for different scales. Weak effect going from corn to ethanol is found at low scales with an increasing tendency towards higher scales. Ethanol is the most elastic at scales between approximately 100 and 350 trading days. Scale dependent estimates are represented by a black curve and 95\% confidence intervals are shown using the grey curves.}\label{fig_ethanol}
}
\end{figure}

Third, we focus on elasticity between ethanol and corn prices. Corn is a primary producing factor of ethanol in the USA and as such, its price changes are mirrored into ethanol price. We study a standard elasticity model $\log(P_{E,t})=\alpha+\beta\log(P_{C,t})+u_t$ where $P_{E,t}$ and $P_{C,t}$ are ethanol and corn prices, respectively. The logarithmic specification allows for interpretation of the $\beta$ coefficient as elasticity, i.e. one percent change in corn is accompanied by $\beta\%$ change in ethanol. Fig. \ref{fig_ethanol} presents the results for daily series between January 2007 and March 2014 (1821 observations). Elasticity varies strongly across scales. For low scales, i.e. in a short run, elasticity remains low yet still positive and increases with an increasing scale. The highest sensitivity of ethanol to changes in corn are reported for scales between approximately half a year and year and a half. Therefore, ethanol reacts to corn very strongly in a long run as one percent change in corn is reflected in ethanol price by between 0.5 and 0.8 percent change. Narrow confidence intervals again support the reported results.

\begin{figure}[!htbp]
\begin{center}
\begin{tabular}{c}
\includegraphics[width=80mm]{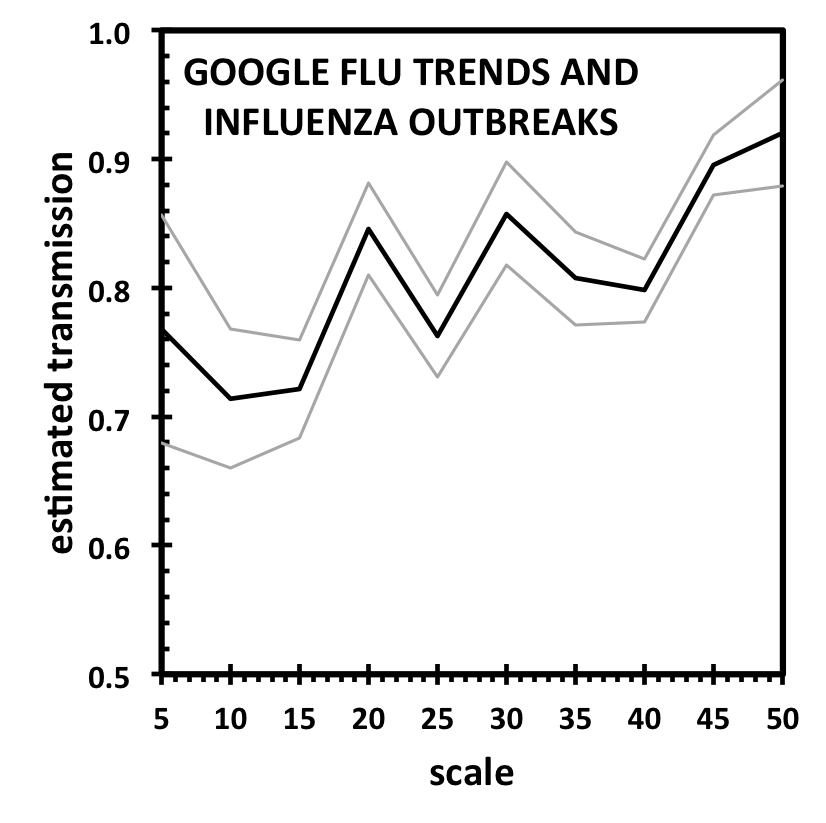}
\end{tabular}
\end{center}
\caption{\textbf{Relationship between Google Flu Trends indicator and real influenza outbreaks.} {The estimated effect varies considerably across scales. Specifically, the informative value of Google Flu Trends indicator increases with the time horizon. From the long-term perspective, the Google Flu Trends indicator reflects the actual influenza outbreaks very well, even though the transmission is not complete, i.e. it is significantly lower than unity. Scale dependent estimates are represented by a black curve and 95\% confidence intervals are shown using the grey curves.}\label{fig_flu}
}
\end{figure}

And fourth, we study the relationship between Google Flu Trends (GFT) indicator and real influenza outbreaks. Ability to anticipate upcoming soars of influenza spreading is an important challenge for scientists from various fields. The GFT indicator has proven its worth in predicting influenza outbreaks even though it has been under some criticism lately \cite{Preis2014}. The basic idea behind the indicator is that people experiencing symptoms of the flu leave a trace online while searching for information about their illness. Google constructs the GFT index using various keywords connected to influenza and their frequency in the search engine usage. Fig. \ref{fig_flu} shows the estimated transmission from GFT to the actual influenza occurrences. We use a weekly dataset provided in a recent study by Preis \& Moat \cite{Preis2014} -- weekly unweighted percentages of patient visits due to influenza-like illness (ILI) provided by Centers for Disease Control and Prevention (CDC) and weekly GFT provided by Google between January 2010 and September 2013 (194 observations). The strength of transmission increases with scale from 0.76 for the lowest scale (5 weeks) to 0.92 for the highest scale (50 weeks). We observe that the confidence bars remain narrow regardless of a relatively limited dataset size in this case. This implies that the GFT indicator contains useful information about the dynamics of influenza occurrence. However, the online-data based index is not able to explain the whole dynamics of the flu spreadings even for higher scales close to a year.

%\section{Conclusion}

In conclusion, we have introduced a new framework of examining relationship between two variables by merging standard regression least squares with detrended fluctuation analysis. The connection not only allows for studying connection between variables at different scales but also provides related standard errors and coefficients of determination. Moreover, the method is constructed to work even for non-stationary and long-range correlated data. The detrended fluctuation analysis regression opens a new area of research in various empirically oriented disciplines and it also delivers a complete framework not necessarily restricted to DFA itself. Other methods for studying long-range correlations such as detrending moving averages, height correlation analysis and others can be easily implemented into the framework as well.\\

%\section*{Acknowledgements}
%The author would like to thank the anonymous referees for valuable comments and suggestions which helped to improve the paper significantly. 

%The research leading to these results has received funding from the European Union's Seventh Framework Programme (FP7/2007-2013) under grant agreement No. FP7-SSH-612955 (FinMaP). 
%Support from the Czech Science Foundation under project No. 14-11402P is gratefully acknowledged.

The research leading to these results has received funding from the Czech Science Foundation project No. 14-11402P and the Research Councils UK via Grant EP/K039830/1.
%\newpage

%\section*{References}
\bibliography{Bibliography}
\bibliographystyle{unsrt}

\newpage

\end{document}